\documentclass[twocolumn]{aastex62}
\usepackage{multirow}

\newcommand{\cyc}{$c-$C$_3$H$_2$}

\graphicspath{{./}{figures/}}

\shorttitle{\cyc\ in TW Hya}
\shortauthors{Cleeves et al.}

\begin{document}

\title{The TW Hya Rosetta Stone Project IV: A hydrocarbon rich disk atmosphere}

\correspondingauthor{L. Ilsedore Cleeves}
\email{lic3f@virginia.edu}

\author[0000-0003-2076-8001]{L. Ilsedore Cleeves}
\affil{University of Virginia, Charlottesville, VA}

\author[0000-0002-8932-1219]{Ryan A. Loomis}
\affil{National Radio Astronomy Observatory, Charlottesville, VA}

\author[0000-0003-1534-5186]{Richard Teague}
\affil{Center for Astrophysics $\vert$ Harvard \& Smithsonian, Cambridge, MA}

\author[0000-0003-4179-6394]{Edwin A. Bergin}
\affil{University of Michigan, Ann Arbor, MI}

\author[0000-0003-1526-7587]{David J. Wilner}
\affiliation{Center for Astrophysics $\vert$ Harvard \& Smithsonian, 60 Garden Street, Cambridge, MA 02138, USA}

\author[0000-0002-8716-0482]{Jennifer B. Bergner}
\altaffiliation{NHFP Sagan Fellow}
\affiliation{University of Chicago, Department of the Geophysical Sciences, Chicago, IL 60637, USA}

\author[0000-0003-0787-1610]{Geoffrey A. Blake}
\affiliation{Division of Chemistry \& Chemical Engineering, California Institute of Technology, Pasadena CA 91125, USA}
\affiliation{Division of Geological \& Planetary Sciences, California Institute of Technology, Pasadena CA 91125, USA}

\author[0000-0002-0150-0125]{Jenny K. Calahan}
\affiliation{Department of Astronomy, University of Michigan, 1085 South University Avenue, Ann Arbor, MI 48109, USA}

\author[0000-0002-1917-7370]{Paolo Cazzoletti}
\affiliation{Leiden Observatory, Leiden University, PO Box 9513, 2300 RA Leiden, The Netherlands}

\author[0000-0001-7591-1907]{Ewine F. van Dishoeck}
\affiliation{Leiden Observatory, Leiden University, PO Box 9513, 2300 RA Leiden, The Netherlands}
\affiliation{Max-Planck-Institut f{\"u}r Extraterrestrische Physik, Giessenbachstra{\ss}e 1, D-85748 Garching bei M{\"u}nchen, Germany}

\author[0000-0003-4784-3040]{Viviana V. Guzm\'an}
\affiliation{Instituto de Astrofísica, Ponticia Universidad Católica de Chile, Av. Vicuña Mackenna 4860, 7820436 Macul, Santiago, Chile}

\author[0000-0001-5217-537X]{Michiel R. Hogerheijde}
\affiliation{Leiden Observatory, Leiden University, PO Box 9513, 2300 RA Leiden, The Netherlands}
\affiliation{Anton Pannekoek Institute for Astronomy, University of Amsterdam, Science Park 904, 1098 XH, Amsterdam, The Netherlands}

\author[0000-0001-6947-6072]{Jane Huang}
\altaffiliation{NHFP Sagan Fellow}
\affiliation{Center for Astrophysics $\vert$ Harvard \& Smithsonian, 60 Garden Street, Cambridge, MA 02138, USA}
\affiliation{Department of Astronomy, University of Michigan, 1085 South University Avenue, Ann Arbor, MI 48109, USA}

\author[0000-0003-0065-7267]{Mihkel Kama}
\affiliation{Department of Physics and Astronomy, University College London, Gower Street, London, WC1E 6BT, UK}
\affiliation{Tartu Observatory, University of Tartu, 61602 T\~{o}ravere, Estonia}

\author[0000-0001-8798-1347]{Karin I. {\"O}berg}
\affiliation{Center for Astrophysics $\vert$ Harvard \& Smithsonian, 60 Garden Street, Cambridge, MA 02138, USA}

\author[0000-0001-8642-1786]{Chunhua Qi}
\affiliation{Center for Astrophysics $\vert$ Harvard \& Smithsonian, 60 Garden Street, Cambridge, MA 02138, USA}

\author[0000-0002-3800-9639]{Jeroen Terwisscha van Scheltinga}
\affiliation{Laboratory for Astrophysics, Leiden Observatory, Leiden University, PO Box 9513, 2300 RA Leiden, The Netherlands}
\affiliation{Leiden Observatory, Leiden University, PO Box 9513, 2300 RA Leiden, The Netherlands}

\author[0000-0001-6078-786X]{Catherine Walsh}
\affiliation{School of Physics and Astronomy, University of Leeds, Leeds LS2 9JT, UK}

\begin{abstract}
Connecting the composition of planet-forming disks with that of gas giant exoplanet atmospheres, in particular through C/O ratios, is one of the key goals of disk chemistry. Small hydrocarbons like $\rm C_2H$ and $\rm C_3H_2$ have been identified as tracers of C/O, as they form abundantly under high C/O conditions. We present resolved \cyc\ observations from the TW Hya Rosetta Stone Project, a program designed to map the chemistry of common molecules at $15-20$ au resolution in the TW Hya disk. Augmented by archival data, these observations comprise the most extensive multi-line set for disks of both ortho and para spin isomers spanning a wide range of energies, $E_u=29-97$ K. We find the ortho-to-para ratio of \cyc\ is consistent with 3 throughout extent of the emission, and the total abundance of both \cyc\ isomers is $(7.5-10)\times10^{-11}$ per H atom, or $1-10$\% of the previously published $\rm C_2H$ abundance in the same source. We find \cyc\ comes from a layer near the surface that extends no deeper than $z/r=0.25$. Our observations are consistent with substantial radial variation in gas-phase C/O in TW Hya, with a sharp increase outside $\sim30$~au. Even if we are not directly tracing the midplane, if planets accrete from the surface via, e.g., meridonial flows, then such a change should be imprinted on forming planets. Perhaps interestingly, the HR~8799 planetary system also shows an increasing gradient in its giant planets' atmospheric C/O ratios. While these stars are quite different, hydrocarbon rings in disks are common, and therefore our results are  consistent with the young planets of HR~8799 still bearing the imprint of their parent disk's volatile chemistry. 
\end{abstract}


\section{Introduction \label{sec:intro}}

We are entering an era where measurements of the compositions of giant exoplanet atmospheres are becoming increasingly common. A diversity of chemical properties (via carbon-to-oxygen ratios, or C/O) has been seen \citep[e.g.,][]{Madhusudhan11,Madhusudhan12,Kreidberg15}, including within a single planetary system \citep[HR8799;][]{bonnefoy2016,lavie2017,lacour2019,molliere2020}. To understand the origins of this diversity, we must study planets' formation environments: gas rich protoplanetary disks. The chemical properties of these disks are driven by at least two factors, i. the make-up of the molecular cloud out of which the star and disk formed \citep[e.g.,][]{visser09,visser11,maria2019}, 
and ii. the disk physical properties (irradiation level, temperature, density, etc.) that can drive an actively evolving chemistry prior to and during planet formation \citep[e.g.,][]{Cleeves14}. 

The relative contribution of these factors, i.e., the role of inheritance versus later chemical reprocessing in the disk itself,  changes with both radial and vertical location. For example, near the highly irradiated disk surface and/or close to the central star, the chemistry is effectively ``scrambled'' leaving little memory of the molecular composition of the cloud. Near the midplane, especially in the outer disk (beyond $\sim10$ au), the high extinction levels provide a safer haven for some of the material originating in the molecular cloud to be preserved, with further processing requiring moderate to high external irradiation or cosmic ray fluxes \citep[e.g.,][]{bergin2014,Cleeves14,yu2017,Eistrup18}; however, it is unclear whether disks are sufficiently ionized to facilitate an active midplane chemistry \citep{Cleeves15}.

There is a growing body of evidence that the observable chemistry in planet forming disks, at least that within the ``warm molecular layer'' \citep{Aikawa02}, deviates from ``typical'' molecular cloud chemistry. 
For example, observations of CO emission and CO isotopologues are faint when compared to expectations based on dust masses from millimeter emission, a gas-to-dust correction factor, and an interstellar CO abundance \citep[e.g.,][]{Favre15,Schwarz16,ansdell2016}. Water was surprisingly challenging to detect in protoplanetary disks with {\em Herschel}, and when detected, fluxes were more than an order of magnitude lower than anticipated based on astrochemical modeling with UV irradiated water ice at interstellar abundances \citep{Bergin10,Hogerheijde11,Du17}. The abundant interstellar complex organic methanol, first detected in disks by the Atacama Large Millimeter Array (ALMA), was similarly quite faint \citep{walsh14}, yet observations of CH$_3$CN appeared relatively bright, with nearly a 1:1 abundance ratio inferred between them for the TW Hya disk \citep{Loomis18b}.

Therefore we need better constraints on the chemical composition of disk gas, especially at an elemental level, to understand the chemical reservoir from which planets may accrete and what, e.g., C/O or N/O ratio they might inherit at least initially. 
Chemical models have demonstrated that the abundances of simple hydrocarbons like C$_2$H and \cyc\ are very sensitive to the C/O ratio of the gas \citep[e.g.,][]{Bergin16,kama2016,cleeves2018,miotello2019,fedele2020}. Pure freeze-out of solids from gas can vary C/O in the gas or ice from $\sim0.2$ to 1 \citep{Oberg11e} around the conventional solar value of 0.54. \citet{cleeves2018} found that the abundance of C$_2$H was extremely sensitive to the bulk C/O in volatiles until C/O $>1$ (ratios of 1.9 and 3.7 were indistinguishable in their models; see also \citet{bosman2021}). Therefore, hydrocarbons are expected to be a useful tool for estimating C/O up to $\sim 1.5$, which overlaps with observed exoplanet atmospheric values.  As a result, disk hydrocarbon studies open an exciting potential avenue to connect the composition of the gas in disks to that measured in planets' atmospheres. In addition, observations of hydrocarbon emission in disks have been found to be generally quite bright at sub-mm wavelengths \citep{Qi13b,Kastner14,kastner2015,Bergin16,cleeves2018,bergner2019}, enabling small surveys ($\lesssim 14$ sources) of C$_2$H emission in disks \citep{guilloteau16,miotello2019,bergner2019}.  

The next step is localizing the distribution of hydrocarbons in the disk, to understand the range of possible C/O values a planet could inherit from a single disk environment. The first resolved image of C$_2$H was made of the TW Hya disk by \citet{kastner2015} with the Submillimeter Array, where it was found to have a ring-like morphology. Its disk-averaged physical nature was constrained using multiple transitions and its hyperfine structure. However, due to the face-on nature of this disk $5-7^\circ$ \citep{Qi06,Qi08,Huang18}, degenerate excitation solutions were found to fit the data, either a cold, dense solution or alternatively a relatively warmer, but low density solution. The former would suggest an enhanced (greater than solar) C/O ratio fairly deep into the disk gas, near the planet-forming region, 
while the latter would suggest that the C/O enhancement primarily is closer to the surface and perhaps less connected to the planet forming midplane. 

\citet{Bergin16} presented further resolved observations confirming C$_2$H's ring-like geometry, as well as observations of \cyc\ toward TW Hya. The spatial distribution of C$_2$H and \cyc\ were found to match identically in radial distribution, suggestive that this radial region of the TW Hya disk supports a generally rich hydrocarbon chemistry, consistent with an elevated C/O ratio in this region.

The present paper uses multi-line observations of \cyc\ to better spatially constrain the nature of the hydrocarbon layer in the TW Hya protoplanetary disk, with the goal of improving our interpretation of the C/O ratio(s) of this disk. The observations were conducted as part of the TW Hya as a Chemical Rosetta Stone Project (PI: Cleeves), and have been augmented with ALMA archival data. The set of lines covers energies spanning 29.1 K to 96.5 K, and with the relatively highly critical density and range of opacities probed, these lines are well suited to constraining the nature, and crucially the location, of small hydrocarbon chemistry in TW Hya. 

\section{Observations\label{sec:obs}}

\subsection{ALMA Observations}

\begin{deluxetable*}{lcccccccc}
\tablecaption{\cyc\ Observations \label{tab:linedat}}
\setlength{\tabcolsep}{5pt} 
\renewcommand{\arraystretch}{.8} 
\tablehead{
\colhead{Line}   & \colhead{o or p} & \colhead{$\nu$ }    & \colhead{$A_{ij}$\tablenotemark{a}} &
\colhead{E$_{\rm u}$\tablenotemark{a}}    & \colhead{{ Channel width}}   & \colhead{rms\tablenotemark{b}} & \colhead{Int. flux\tablenotemark{c} } & \colhead{Program\tablenotemark{d}} \\
 & & {(GHz)} &  (s$^{-1}$) &{(K)}    & (km/s)  & { (mJy/beam)} & { (mJy km/s)} & 
}
\startdata
$4_{3,2} - 3_{2,1}$ & o & 227.16913 & 3.113E-04 & 29.07 & 0.16 & 3.2 & $80 \pm 5$ & 1  \\
$7_{3 , 4} - 6_{4 , 3}$ & o & 351.52327 & 1.237E-03 & 77.24 & 0.21 & 3.4 & $237 \pm 13$ & 2 \\
$8_{3 , 6} - 7_{2 , 5}$ & o & 352.19364 & 1.734E-03 & 86.93 & 0.21 & 4.2 & $220 \pm  30$ & 3  \\
$5_{5  , 1} - 4_{4 , 0}$ & p & 338.20399 & 1.598E-03 & 48.78 & 0.21 & 5.3 & $140 \pm 16$ & 3 \\
$8_{2 , 6} - 7_{3 , 5}$ & p & 352.18551 & 1.735E-03 & 86.93 & 0.21 & 4.3 &  $96 \pm 15$ & 3  \\
$9_{1,8}   - 8_{2,  7}$ & o (bl) & 351.96593 & 2.117E-03 & 93.34 & \multirow{2}{*}{0.21} &  \multirow{2}{*}{2.9} &  \multirow{2}{*}{$398 \pm 14 $} & \multirow{2}{*}{2,3}  \\
$9_{2,8}-8_{1,7}$       & p (bl) & 351.96594 & 2.117E-03 & 93.34  &  & & &  \\
$10_{1,10} - 9_{0,  9}$ & o (bl) & 351.78157 & 2.439E-03 & 96.50  & \multirow{2}{*}{0.21} &  \multirow{2}{*}{2.9}  & \multirow{2}{*}{$ 330 \pm 15$} & \multirow{2}{*}{2,3}  \\
$10_{0,10}-9_{1,9}$ & p (bl) & 351.78157 & 2.439E-03 & 96.50  & &  & &   
\enddata
\tablenotetext{a}{Line catalogue data from CDMS \citep{Muller05}}
\tablenotetext{b}{In a $0\farcs5$ beam, with the specified channel width}
\tablenotetext{c}{Measured within a Keplerian mask (see Section~\ref{sec:methods}, and Figure~\ref{fig:channels}).}
\tablenotetext{d}{1) 2013.1.00114.S, 2) 2016.1.00311.S, 3) 2013.1.00198.S }
\end{deluxetable*}

The observations of \cyc\ utilized in the present study were carried out with ALMA as part of three different observational programs. We present new observations of \cyc\ taken as part of the TW Hya as a Chemical Rosetta Stone Program (PI: Cleeves, 2016.1.00311.S), augmented by archival observations from 2013.1.00198.S (PI: Bergin) and 2013.1.00114.S (PI: {\"O}berg). The complete set of \cyc\ transitions from these programs used in our analysis are listed in Table~\ref{tab:linedat}. 
The observations from 2016.1.00311.S were carried out with 40 antennas on April 8, 2017 (C40-2; 15m -- 390m baselines) and May 21, 2017 (C40-5; 15m -- 1124m baselines). The April 8 observation used J1037-2934 for bandpass and phase calibration, and J1058+0133 for flux calibration. The May 21 observation used J1037-2934 for bandpass,  flux, and phase calibration.

\begin{figure*}
\begin{centering}
\includegraphics[width=0.83\textwidth]{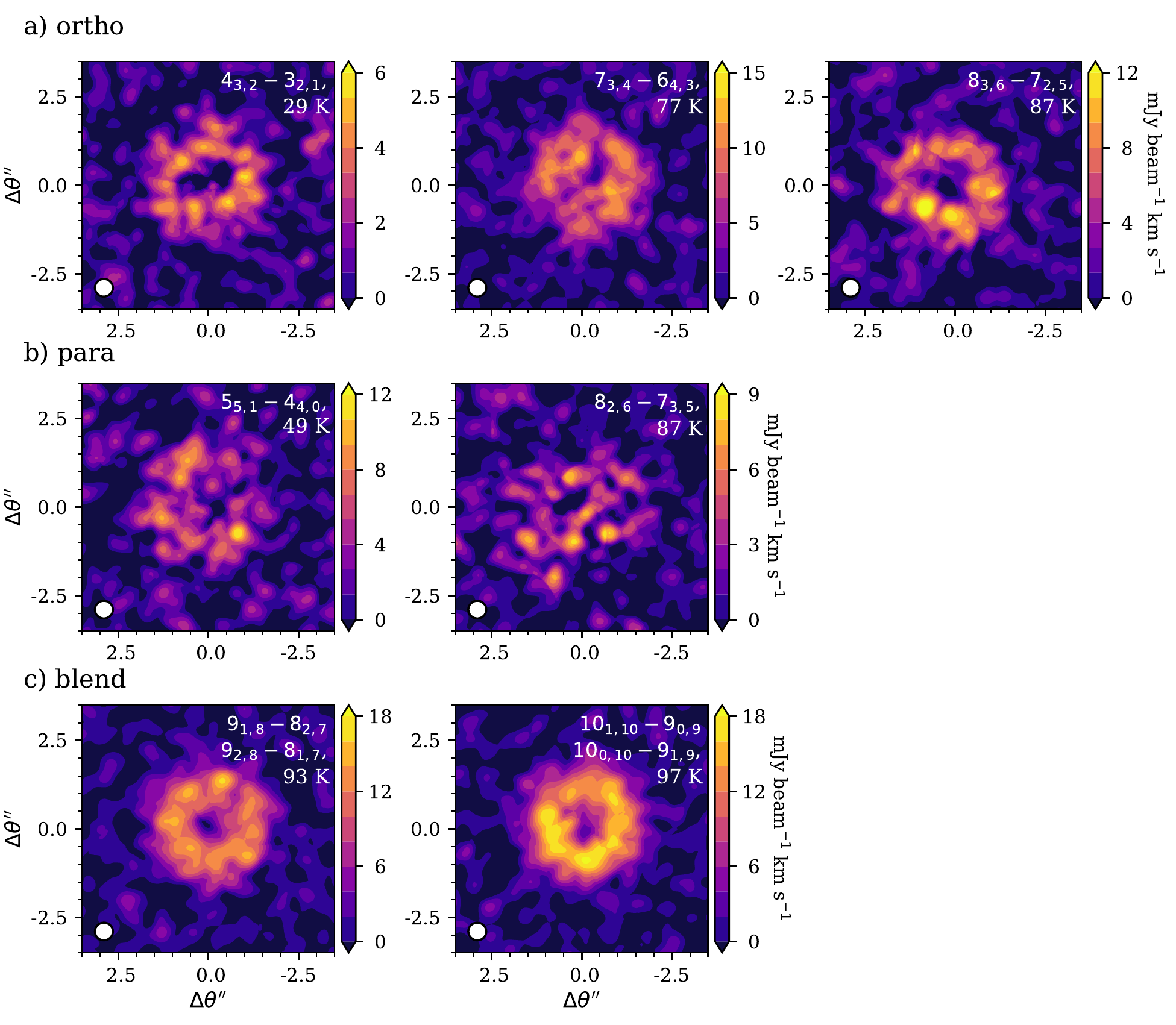}
\caption{Moment zero maps for the five isolated transitions of a) ortho, b) para, and c) two blended transitions of \cyc\ that are analyzed in this work. Line ID and upper state energy are labelled in the upper right corner of each panel.  \label{fig:mom0}}
\end{centering}
\end{figure*} 

All data were calibrated with the ALMA pipeline with CASA version 4.7.0. Prior to imaging, we phase self-calibrated the data using the line-free portions of the continuum, adopting a solution interval of 30s and averaging polarizations. In addition, spws were self calibrated independently, and we set a minimum signal to noise ratio of 3 and a minimum number of baselines per antenna of 6. 
The calibration process for the observations from program 2013.1.00198.S and 2013.1.00114.S are reported in \citet{Bergin16} and \citet{Oberg17} and not repeated here.

The projects 2013.1.00198.S and 2016.1.00311.S had overlapping coverage of the bright 10 -- 9 and 9 -- 8 blends (see Table~\ref{tab:linedat}). The integrated line flux measured between these two programs differed by about 10\% (well within the quoted ALMA flux uncertainty), however we adjusted the flux of 2013.1.00198.S to match the flux measured in 2016.1.00311.S flux. As a result, 6 of the 7 lines have internally consistent fluxes, and therefore RMS uncertainties reported here do not include flux calibration uncertainty, as that will either net increase/decrease measured flux but will not impact the spatially resolved shape of the line images.

\subsection{Images and Radial Integrated Flux Profiles} 
The continuum was subtracted in the uv-plane using the CASA task  \texttt{uvcontsub} assuming a linear fit to the continuum shape. Imaging was carried out using the \texttt{tclean} task in a semi-automated fashion described here. The source velocity was assumed to be 2.84 km s$^{-1}$. For each line, using all available data for a given transition, we begin by creating a dirty image to determine the standard deviation per beam in line free channels and the beam size for each transition. From these data, we create a mask based on expectations from Keplerian motion using the code presented as part of \citet{Pegues2020}, see also code reference \citet{jpegues2020code}. We assume an inclination of $5^\circ$, a position angle of $152^\circ$, and a stellar mass of 0.8~M$_\odot$ to create the Keplerian mask following \citet{Huang18}. We assume a distance of 59.5~pc \citep{GAIA16}. The mask is convolved with the respective beam for each data set (typically between $0\farcs2 - 0\farcs4$). Because we want to have a consistent beam size across all the transitions, which we chose to be $0\farcs5$, we calculate the uv-taper that is necessary to create a beam of just below $0\farcs5$ on the major axis. We then clean each line with its respective mask in non-interactive mode down to a noise level of 3$\sigma$ standard deviation, with the line specific uv-taper applied. Finally, we take the cleaned image and apply the CASA task imsmooth, which operates in the image plane, to create a final image with an exactly $0\farcs5$ round beam.   The result of this imaging process is shown in Figure~\ref{fig:mom0}. 
The corresponding channel maps for all transitions are also provided in Appendix~A, Figure~\ref{fig:channels}.

\begin{figure*}[t]
\begin{centering}
\includegraphics[width=0.93\textwidth]{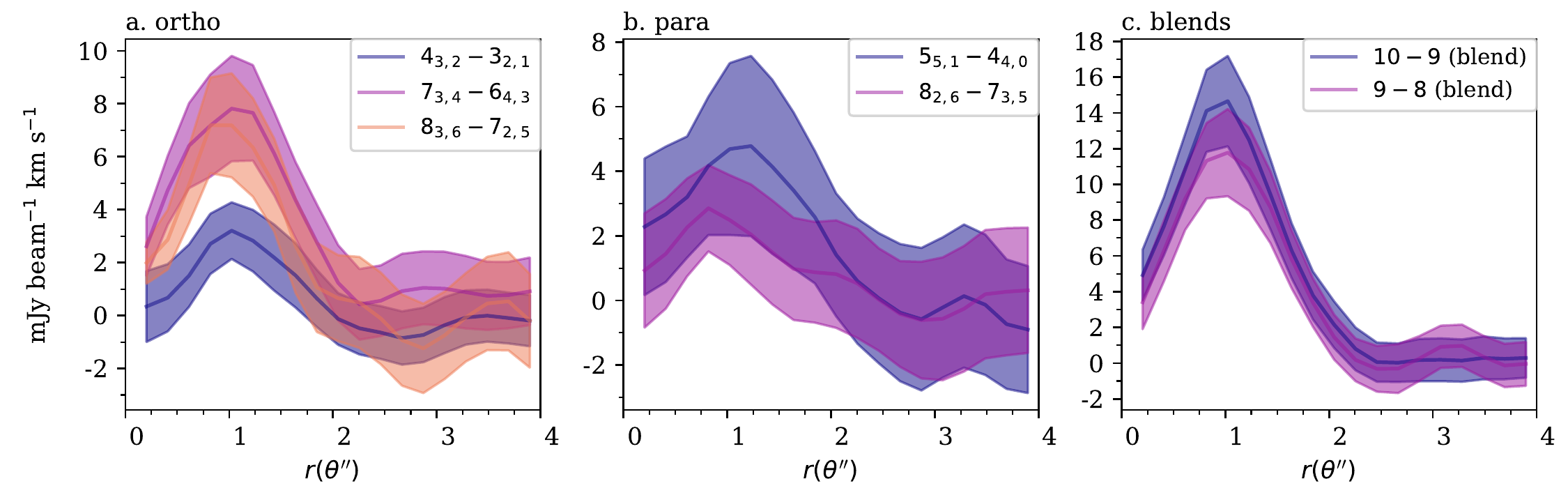}
\caption{Profiles extracted by the pixel stacking technique within the GoFish code \citep{gofish}. Transitions indicated in the legend. Shaded region indicates the error on the radial profile. \label{fig:gofishprof}}
\end{centering}
\end{figure*}

The ring-like nature of the emission is clear from the moment-0 maps, and the general faintness of the para transitions compared to the ortho transitions is also clear. We do not use these moment-0 maps for analysis, showing them here to illustrate the detection significance, and instead we apply the \textsc{GoFish} package \citep{gofish} to improve the signal to noise of the radial profile for each transition. \textsc{GoFish} works in the image plane to spectrally align spectra at a given radius by taking into account the projected disk rotation, such that the spectra can be stacked, thereby creating a higher signal-to-noise ratio radial intensity profile. We assume the same disk parameters as for the Keplerian masking above to carry out the \textsc{GoFish} deprojection. The deprojection is done on the cleaned image cubes produced with the method described above, and the output is a spectrum at each radius where the width of the emission is some combination of the thermal width and any non-thermal broadening or beam convolution broadening, since Keplerian motion has been accounted for. After shifting and stacking the azimuthal emission, we integrate over a conservative velocity range of 2.5 km~s$^{-1}$ to 3.18 km~s$^{-1}$ and obtain the profiles shown in Figure~\ref{fig:gofishprof}. Note that even though the para transitions are faint in the moment-0 maps in Figure~\ref{fig:mom0}, the ring-like structure becomes clearer in the \textsc{GoFish} extraction and the peak of the resolved profile is detected at $\ge 3 - 4 \sigma$.

\section{Methods \label{sec:methods}}

From this multi-line data set of \cyc\ rotational transitions spanning upper state energies from 29.1 K to 96.5 K, we aim to use excitation to constrain the location of the hydrocarbon layer in the TW Hya disk. To fit the data, we use a simple slab model method \citep[see e.g.,][]{Qi08,Oberg17} and employ a non-LTE line radiative transfer code based upon Ratran \citep[][Astrophysics Source Code Library, record 0008.002]{hogerheijde2000} designed to handle blended transitions and be computationally efficient (https://github.com/ryanaloomis/nLTErt1d). The inputs to this code are height ($z$), gas volumetric density, dust density, gas temperature, dust temperature, non-thermal line width, and abundance relative to hydrogen of the molecule of interest. 

We approximate the emission from TW Hya as a series of radial annular regions, where the physical conditions as a function of height at each radial location are taken from the \citet{Cleeves15} model of TW Hya. The gas temperature, dust temperature, and gas density are taken from this model, where the total mass of this TW Hya disk model is 0.04 $M_\odot$.

\begin{figure*}
\begin{centering}
\includegraphics[width=.95\textwidth]{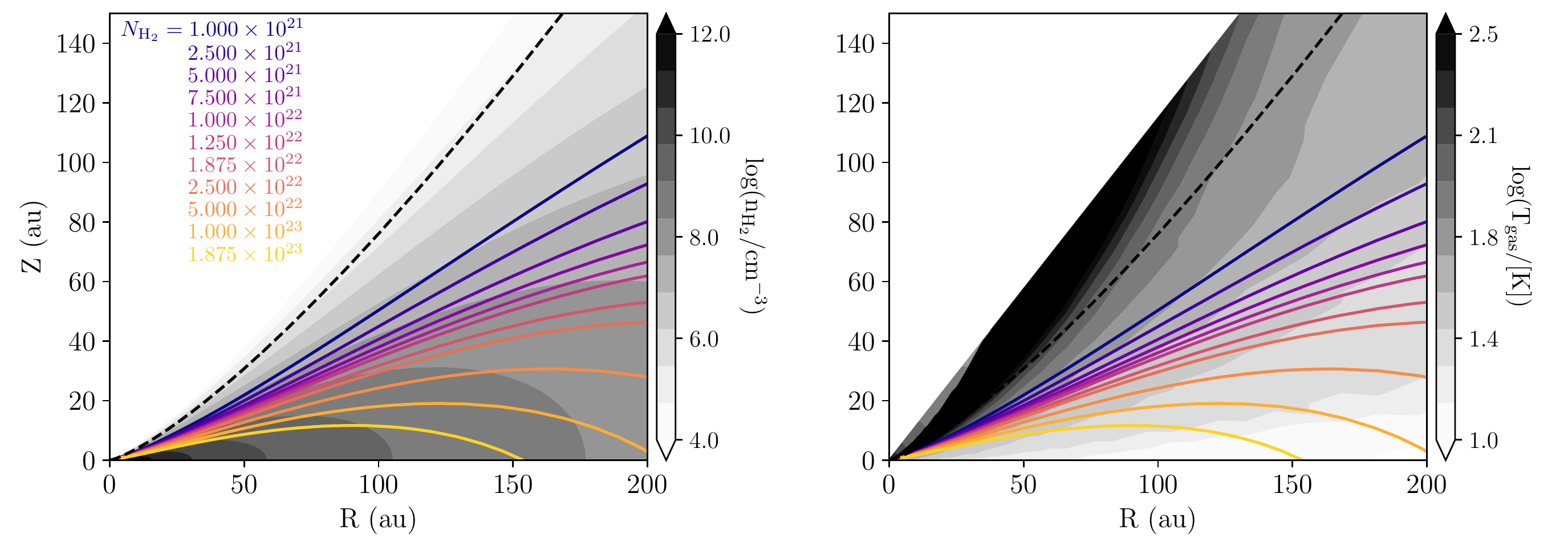}
\caption{Slab \cyc\ model definitions. The vertical extent of \cyc\ is described by the vertically integrated column density derived from the \citet{Cleeves15} TW Hya model shown in grey shaded contours. The upper limit for all models is $N_{\rm H_2}=10^{19}$ cm$^{-2}$ (dashed line). The bottom of the \cyc\ slab is described by the color line contours as labelled in the left panel. The inner and outer radius parameters truncate these boundaries vertically, and the abundance is assumed to be constant inside of the layer. \label{fig:slabcartoon}}
\end{centering}
\end{figure*}

The ``slab'' \cyc\ distribution is bounded by an inner and outer radius, and a vertically computed upper and lower column density of H, $N_{\rm H_2}$. Note that for the ISM, $N_{\rm H_2} = 1.9\times10^{21}$ cm$^{-2} = 1~A_{V}$. But since our model has 6.7$\times$ less small grains in the surface of the disk due to the formation and subsequent settling of larger grains, then $1~ A_{ V}$ occurs at $N_{\rm H_2} = 1.27\times10^{22}$ cm$^{-2}$.  The general reasoning for this vertical parameterization rather than a simple $z/r$ cut is that we expect the \cyc\ distribution to be driven by the stellar radiation field, largely UV \citep{Du15,Bergin16}. The abundance is set to a constant value within this ``slab,'' where we vary the value of this constant.  The underlying physical structure and the calculated $N_{\rm H_2}$ contours bounding the \cyc\ distribution are shown in Figure~\ref{fig:slabcartoon}. As can be seen in the right-hand panel, the slabs closer to the disk surface are dominated by warmer temperatures, while deeper in (higher $N_{\rm H_2}$) the gas temperatures decrease.

In addition to the four parameters described above, we also fit for the ortho to para ratio of \cyc, where as can be seen in Table~\ref{tab:linedat} our data set contains both spin isomers and two blended pairs of ortho and para transitions (the brightest lines). We adopt the collisional rate coefficients hosted on the Leiden LAMDA database \citep{Schoier05} originally computed in \citet{chandra2000} for ortho and para separately. For the blended transitions, the components of the blends had identical molecular parameters (e.g., Einstein A coefficients, near identical frequencies), and therefore we simulated emission for the blends by first estimating the level populations for each of the ortho and para components independently (while we did not impose it on the solution, the emission was seen to be largely in LTE). We then took a weighted average of the ortho and para populations in the upper and lower states respectively (the population fractions were the same for both spin isomers). This allowed us to simulate the blended \cyc\ emission as that from a ``single'' molecule. 
We took this approach since for some of the models the blended transitions became optically thick, so we could not simply sum the line fluxes from the independent components without potentially overestimating the blended line flux.

\begin{deluxetable}{lcc}
\tablecaption{Slab Model Parameters \label{tab:params}}
\setlength{\tabcolsep}{5pt} 
\renewcommand{\arraystretch}{.8} 
\tablehead{
\colhead{Parameter}   & \colhead{Values} & \colhead{Unit}
}
\startdata
Min H$_2$ Column & $10^{19}$ & cm$^{-2}$ \\
Max H$_2$  Column ($\times 10^{21}$) & 1, 2.5, 5, 7.5, & cm$^{-2}$\\
      & 10, 12.5, 18.75, 25,  &\\
                        & 50, 100, 187.5 &\\                        
$\chi$(\cyc)   ($\times 10^{-11}$)& 1, 1.875, 3.75, 5, & per H\\
      & 7.5, 10, 25, 50, 75, &\\
                        &  100, 250, 500, $10^3$ & \\
OPR & 1, 3 & n/a \\
$R_{\rm inner}$ \cyc\ & 20, 25, 30, 35, 40 & au\\
$R_{\rm outer}$ \cyc\ & 80, 90, 100, 110, 120 & au\\
Total number of models & 7150 & 
\enddata
\end{deluxetable}

To create the synthetic integrated radial intensity profile for each model, we reconstruct a 2D sky image of the velocity integrated line brightness and convolve the image with a Gaussian $0\farcs5$ beam. The use of the larger beam was primarily to be able to circularize and homogenize the beam for all lines in the sample and to improve the signal to noise per beam. We then extract the convolved radial brightness profile in annuli with the same spacing as the \textsc{GoFish} intensity profile.

Goodness of fit is assessed based on a reduced $\chi^2$ between the beam-convolved model spectrally integrated intensity profile and the data. To estimate $\chi^2$, the flux in a given annulus for the data and model are measured per beam, and we sum the square of the difference divided by the observed RMS uncertainty.  The number of annuli super samples the beam but is divided out by the reduced $\chi^2$. All lines are treated equally in our overall assessment of fit for a given set of slab model parameters, even though some lines are brighter than others. We made this decision since the brighter lines are optically thick blended transitions, and the lower SNR profiles provide important constraints on our fit.  
Note, we only consider the RMS uncertainty in this estimate and do not include flux calibration error since eight out of nine lines have self-consistent fluxes, such that flux uncertainty  will globally shift all radial line profiles upwards or downwards rather than expanding the error bars uniformly. The only exception is the ortho \cyc\ line $4_{3,2} - 3_{2,1}$.  We therefore  have also examined our best fit models from the profiles as extracted directly from \textsc{GoFish}, and then with a 10\% uniform increase and decrease on just the $4_{3,2} - 3_{2,1}$, with the rest of the lines fixed. We do not find a change in the best fit models, in part since we have two other ortho lines. Furthermore, the models that fit the rest of the data give reasonable fits to the native $4_{3,2} - 3_{2,1}$ observations without flux scaling, and therefore we do not vary the flux for the remainder of the analysis described here.

Initially, we also varied the minimum column density of the layer, but since we did not see a significant change between changing the upper boundary from $N_{\rm H_2} = 1\times10^{19}$ cm$^{-2}$ to $1\times10^{20}$ cm$^{-2}$, we decided to fix the upper boundary for the full grid of simulations to the former value of $1\times10^{19}$ cm$^{-2}$, which for the reduced dust atmosphere corresponds to an $A_V$ of 0.0008, i.e., very high up in the atmosphere, where the H$_2$ gas density is also very tenuous {($\sim 10^{5}$~cm$^{-3}$)}. Thus very little \cyc\ in this layer contributes to the emission. The full set of parameters and their values considered in the simulation grid are provided in Table~\ref{tab:params}. Essentially, each model varies the inner and outer extent, lower layer extent via the column density, the abundance within the layer, and the ortho-to-para ratio of \cyc. The range of values was explored is based on a combination of previous modeling results \citep[e.g.,][]{Bergin16} along with making sure we explored a wide space around regions that gave better fits to assess degeneracies in the parameter space. 

\section{Results \label{sec:results}}

\begin{figure*}
\begin{centering}
\includegraphics[width=1\textwidth]{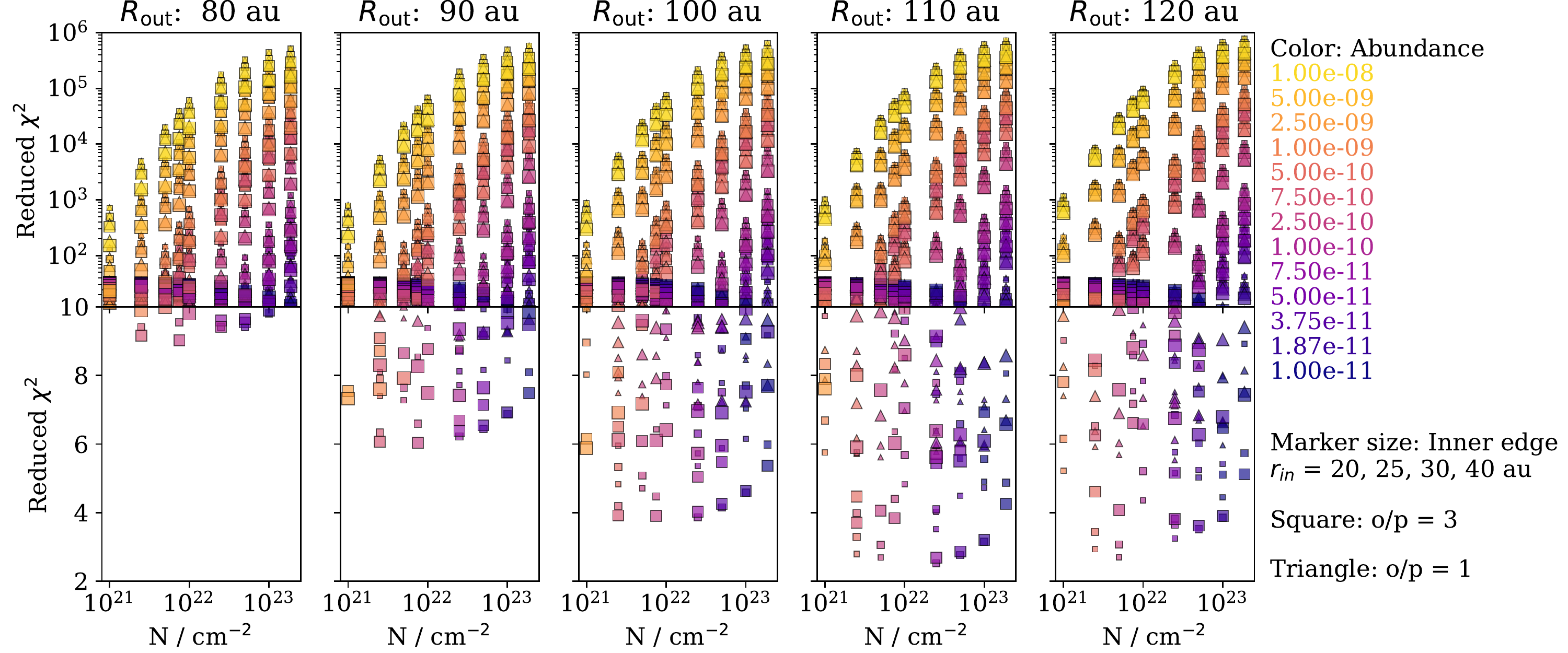}
\caption{Reduced $\chi^2$ between the observed visibilities and modelled visibilities. Each point represents the model fit for all observed lines as a summed reduced $\chi^2$. The model parameters are described by the point color (abundance), size (inner radius of \cyc), and shape (ortho to para ratio). The columns indicate different outer radii for the \cyc\ slab, for the full range of H$_2$ column densities considered on each x-axis. The poor fits are shown on a logarithmic scale on the top row, while the better fits are shown as a linear scale on the bottom row. \label{fig:allmods}}
\end{centering}
\end{figure*} 

\subsection{Fiducial Model Results}
A comparison between all of the 7,150 models in our grid and all observed transitions is illustrated in Figure~\ref{fig:allmods}. Each point represents $\chi^2_{\rm reduced}$ for all observed lines simultaneously.  Due to the large dynamic range across the model grid, we have split each panel into a top and bottom panel where the top row has log scale $\chi^2_{\rm reduced}$ and the bottom row shows the same data but with a linear scale for clarity. The best fit models (those with the lowest $\chi^2_{\rm reduced}$) are those that appear on the bottom row. From Figure~\ref{fig:allmods}, a few key features become clear, and are enumerated here:
\begin{enumerate} 
\item The models that agree the best with the data  have the outer edge of the \cyc\ emission at $>100$ au.
\item None of the other parameters in our grid can be tuned to achieve a model with a good fit with an outer edge at $<100$ au. 

\item An inner edge of 25 and 30 au tends to be a better fit for the \cyc\ distribution. We note that this edge is close to the half-beam of the data (0\farcs25 = 15~au), and so cannot constrain this any further from existing data. We also note that the \cyc\ ``empty'' inner disk model is consistent with the non-zero flux near the star seen in our radial profile due to beam convolution effects. Finally, we note that both the value for the inner and outer edge only hold if a slab model is an adequate representation of the data, and certainly more complex structural representations could be explored with higher SNR data.

\item The flatness of the $\chi^2_{\rm reduced}$ for the best models suggests some degeneracy between the maximum column density, i.e., the depth of the layer, and the abundance of \cyc\ in the layer, which is to be expected especially if the emission is thermalized. 

\item While the observations suggest the \cyc\ originates in a region of sufficiently high gas density to be thermalized, we can further constrain the vertical location of the \cyc\ layer by taking advantage of the wide span of upper state energies in our data set. We can rule out \cyc\ emitting from the upper surface (i.e., $N_{\rm H_2} \le 10^{21}$ cm$^{-2}$) or from the midplane, below $N_{\rm H_2} \ge 10^{23}$ cm$^{-2}$. These column densities translate to $A_V\le0.05$ and $A_V\ge 5$, respectively, due to the dust-poor gas in the disk surface. 

\item There are no good fits that have an ortho-to-para ratio of \cyc\ of 1. Ortho-to-para of 3 is strongly favored over 1. We do not try to fit between these values given the signal to noise of the data.
\end{enumerate}

\begin{figure*}
\begin{centering}
\includegraphics[width=1\textwidth]{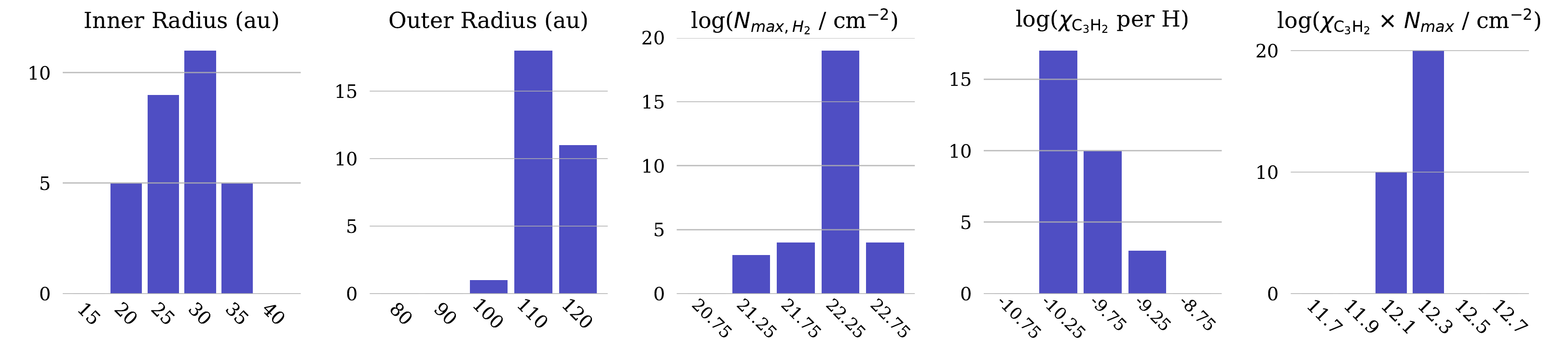}
\caption{Histogram of the parameters of the 30 best fit models. Note 1 $A_{\rm V}$ Corresponds to $N_{\rm H_2} = 1.27 \times 10^{22}$ cm$^{-2}$ given the reduced dust in the upper layers of TW Hya.  \label{fig:histor}}
\end{centering}
\end{figure*}

\begin{figure*}
\begin{centering}
\includegraphics[width=.98\textwidth]{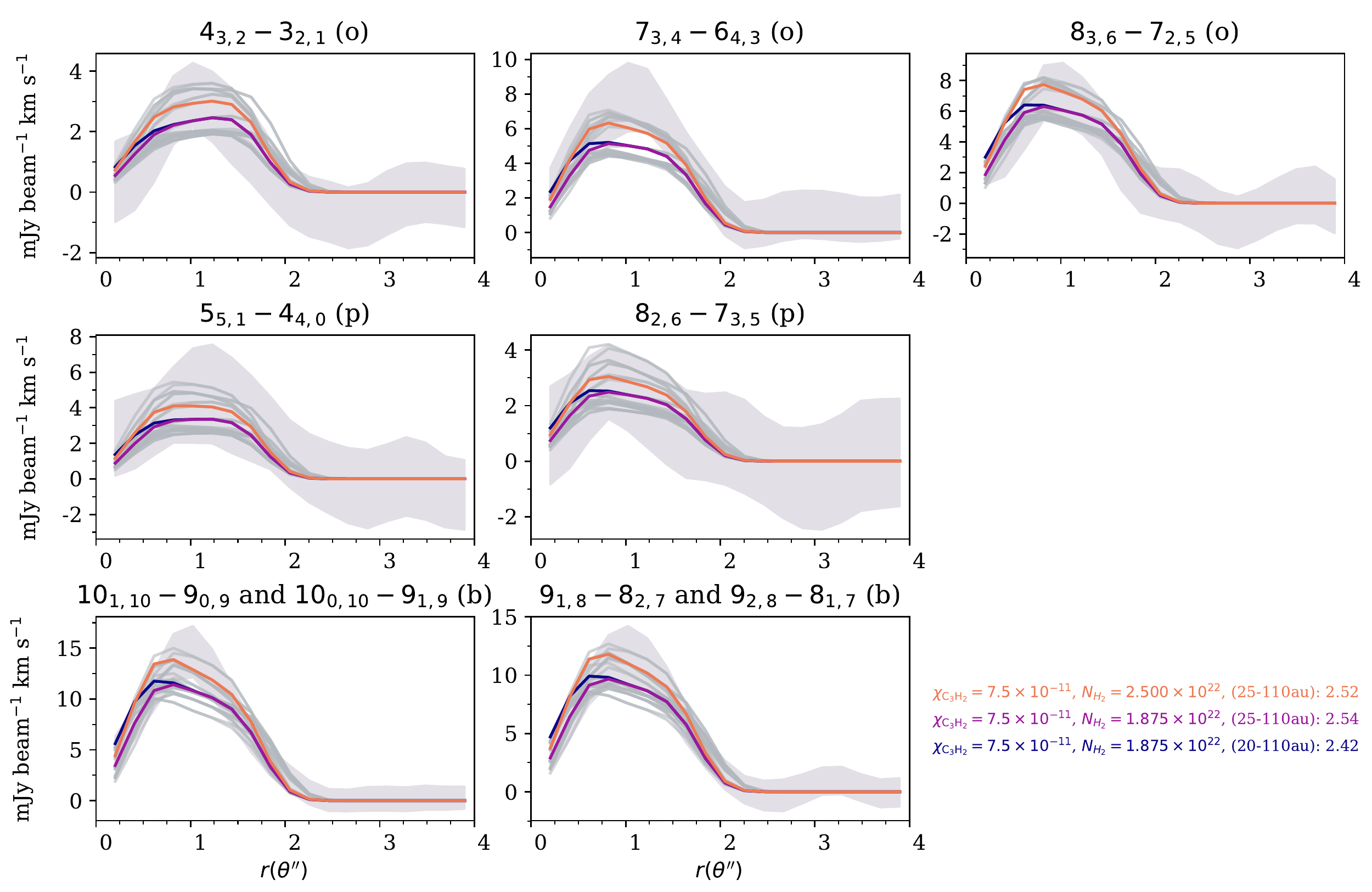}
\caption{Radial profiles of the integrated emission from each transition, including error bars (shaded gray). Profiles extracted using GoFish. The thirty best fit models are show as gray lines, while the top three models are shown in color, and have parameters indicated in the legend in the lower right-hand corner.
\label{fig:proffit}}
\end{centering}
\end{figure*}

To provide an even clearer picture of our best fit models, we extract the 30 models with the lowest $\chi^2$ from our grid and plot a histogram of the model parameters (see Figure~\ref{fig:histor}). We also plot the radial intensity profiles of these models in Figure~\ref{fig:proffit}. Based on this sub-selection of models, an inner radius of 25 or 30 au is favored by 2/3 of the 30 models. Nearly all of the best fit models have an outer radius of \cyc\ at $>100$ au, with 110 au favored slightly over 120 au. While there is some degeneracy between the depth of the slab measured by $N_{\rm max}$ and the abundance per H of \cyc, it is clear that models with $N_{\rm max}$ of $(1-3) \times 10^{22}$ cm$^{-2}$ are favored by most. The abundance of \cyc\ in the slab is between $(3-30) \times 10^{-11}$ per H atom. To test how much of the spread is due to the degeneracy between the thickness of the layer and the abundance, we also show in Figure~\ref{fig:histor} a histogram of the product of $\chi_{\rm c-C_3H_2}$ and $N_{\rm max}$ and find that this distribution is much tighter, and corresponds to an approximate column density of \cyc\ in the slab of $(1-3) \times 10^{12}$ cm$^{-2}$.

The 30 best fit models are shown in Figure~\ref{fig:proffit} in dark grey, and we have additionally highlighted the four best fit models in color and provided their physical parameters in the key. Note all 30 models have an ortho to para ratio of 3, so that parameter is not listed in the figure. The best fit models have similar characteristics, and abundance of $(7.5-10) \times 10^{-11}$ per H atom, and extend down to an H$_2$ column density  of  approximately $2\pm0.5 \times 10^{22}$ cm$^{-2}$. This depth corresponds to a vertically integrated $A_V$ of about 1 due to the reduced small dust mass in the disk upper layers in the \citet{Cleeves15} TW Hya model. 

\section{Discussion \label{sec:disc}}

We have carried out a resolved, multi-line study of hydrocarbon emission in the TW Hya disk as traced by \cyc. The observations span a wide range of upper state energy from 29 K to 96.5 K, and include both ortho and para forms of \cyc. Hydrocarbons like \cyc\ and the more widely studied C$_2$H have been suggested to be good tracers of C/O ratios in disks, which is a parameter that we are now beginning to measure in the atmospheres of gas-rich exoplanets, which should be strongly influenced by (if not set by) the gas in their parent disk \citep[e.g.,][and more]{Oberg11e,Oberg16b}.  Therefore, it is crucial to understand the nature of the gas that these species trace, and specifically {\em where} in the disk they are located to understand how closely it links to what forming planets might accrete. 

The present study augments previous studies of the chemically related species, C$_2$H, also observed toward the TW Hya disk \citep{kastner2015,Bergin16}. We confirm the earlier finding of \citet{Bergin16} that the radial extent of \cyc\ (qualitatively determined by stacking) appears very similar to the brighter C$_2$H emission. If we assume C$_2$H and \cyc\ originate from a similar layer, e.g., in $z/r$ space or column density space, the present study provides further context on the vertical distribution of the hydrocarbon layer in TW Hya, and what that means for eventually using this species to measure C/O in this disk \citep[see, e.g.,][]{cleeves2018}. In the following sections we discuss our main findings.

\subsection{\cyc\ is located in and above the warm molecular layer in a ring}
We find that all models that provide reasonable fits to the full data set do not extend much further below an H$_2$ column density of a few $\times 10^{22}$ cm$^{-2}$. Given that the small dust mass in this model is only 15\% of the interstellar value, this layer corresponds to a maximum $A_V$  of 1 -- 2. We do not have strong constraints on the upper extent of the layer, and where additional observations for lines with higher upper state energies (and smaller Einstein A coefficients) would be helpful. While this lower boundary does not directly correspond to a normalized height value of $z/r$ due to the nature of the assumed density profile (a power-law with an exponential taper), we can say that within the radial boundaries of the emission, that the \cyc\ arises above a $z/r$ value of approximately 0.25 based upon Figure~\ref{fig:slabcartoon}, and no higher than 0.4. This range is consistent with the location of C$_2$H in this source as modeled in \citet{Bergin16}.

We also constrain the radial distribution of the \cyc\ layer. The inner edge of the \cyc\ distribution (25 -- 30 au) is consistent with the CO snow line \citep{Qi13c,Zhang17} and/or the dip in CO emission \citep{Huang18} and scattered light \citep{debes2013,vanBoekel16}. The inner deficit of  \cyc\ might be signalling a large chemical shift, perhaps a large reduction of C/O at this radius. Models have predicted an enhancement of CO interior to the CO snow line due to dust evolutionary processes \citep{krijt2018}; however, CO adds equal parts C and O and will not tend to reduce C/O especially if it is somewhat elevated (above solar, 0.54) but still below 1 already. An alternative explanation is that the same process that sequestered water ice into larger grains has begun in a delayed way on the carbon-bearing molecules \citep[e.g.,][]{cleeves2018}. This process would be fastest in the inner disk and would move outward, which might explain why different disks have a wide variety of C$_2$H ring patterns \citep[e.g.,][]{Bergin16}. Seeing whether these rings have a predictable time evolution for a statistically significant sample of disks would help elucidate the cause \citep[see also discussion in][]{bergner2019}. It would likewise be interesting to compare with other species that also show inner chemical deficits, like CN \citep{cazz18}.

The outer edge of \cyc\ is located at 110 -- 120 au. It is not clear why this radius is remarkable, besides that it corresponds also to the edge of C$_2$H \citep{Bergin16}. The millimeter pebble disk extends to approximately 60 au, however there is a flatter weakly emitting ``shoulder'' in the ALMA observed 852 micron flux out to around 100 au \citep{Andrews12,Andrews16,vanBoekel16,Huang18}. The CO disk extends out beyond 200 au \citep{Andrews12,Schwarz16,Huang18}, though there is a sharp drop in $^{13}$CO at around 100~au \citep{Zhang17}. While the scattered light extends out to 200 au like the CO gas disk, it has some potentially interesting features around 100 au. For example, there is a brightness peak in the observed NIR scattered light at about 100 au \citep{vanBoekel16}. \citet{debes2013} attributes this change at 100 au to be potentially related to sculpting by a Neptune-mass planet. In addition, there appears to be a shift in the ``color'' of the scattering at 100 au, where it is neutral interior to 100 au and becomes more blue outside of 100 au \citep{debes2013,vanBoekel16}. Therefore this location seems to be an important transition; however, its nature remains unclear. Perhaps it marks a significant change in gas density (i.e., H$_2$) or alternatively a chemical change where the disk becomes so tenuous that the external UV field becomes too harsh and densities too low to support chemistry more advanced than CO.

\subsection{The rotational emission of \cyc\ appears to be thermalized}

We find that our best fit models have some spread in $N_{\rm max}$ and $\chi_{\rm c-C_3H_2}$ that would be expected given that we are more fundamentally tracing a column density of \cyc\ with our observations, especially due to the face on nature of this disk. Within these models, most of the emission is coming from the bottom of the layer, and the emission appears well represented by LTE \citep[see also][for a non-LTE analysis of CS]{teague2018}. To provide context, some models with the same \cyc\ column density placed very high up in the disk, well above $A_V = 1$ with a very high \cyc\ abundance and a low value of $N_{\rm max}$ do not provide good fits to the observations. These models are also no longer thermalized.  We conclude that the \cyc\ appears to be reasonably well approximated by LTE. The critical density for the transitions observed is typically around $10^6$ cm$^{-3}$, and our results suggest there remains a substantial amount of H$_2$ gas, $>10^6$ cm$^{-3}$, at high ($z/r > 0.25$) altitudes and moderately far out radii (25 -- 110 au), consistent with disk mass estimates provided by HD for this source \citep{Bergin13, Favre13,Cleeves15,Trapman2019,Calahan20}. How this relatively old disk maintains such a large amount of gas, however, is still unclear.

\subsection{The abundance of \cyc\ relative to C$_2$H is consistent with gas-phase chemical models}

\citet{Bergin16} modeled the related species C$_2$H in TW Hya using full thermo-chemical models. Consistent with the results of \citet{Du15}, in this work it appeared clear that to reproduce the observed brightness of C$_2$H the carbon relative to oxygen ratio of the gas must be very high, greater than the solar ratio of 0.54. To obtain a high C/O ratio, either there must be a source of carbon enhancement, such as carbon grain or PAH destruction \citep{anderson2017}, and/or a deficit of oxygen, possibly due to grain growth and settling \citep{Hogerheijde11,Salinas16,cleeves2018}. The same will be true for \cyc. Taking our best fit model abundances of \cyc\ and comparing it to the results of \citet{Bergin16} for C$_2$H in TW Hya, we find \cyc's abundance is $\sim 3-30\%$ that of C$_2$H. If instead we use the column density, for \citet{Bergin16}'s C/O = 1 model, the \cyc\ to C$_2$H ratio is approximately 10\%. While detailed chemical modeling is beyond the scope of the present paper, we can compare this percentage to a published model of a different disk, IM Lup, where C/O was also varied \citep{cleeves2018} purely by removal of volatile oxygen (via water sequestration into large grains, presumably settled into the midplane as well as O-removal from CO for the more extreme C/O ratios). In the $z/r$ layer of 0.25 -- 0.5, that model finds a \cyc\ to C$_2$H percentage of 1 -- 10\%, broadly consistent with our findings here, without the need for additional carbon sources. In comparing these percentages with other star-forming environments, we see that this ratio of \cyc\ to C$_2$H is also broadly consistent with what is observed in PDRs \citep[e.g., $\sim3$\% toward the Orion Bar][]{Cuadrado15}.

While the chemistry is consistent with gas phase routes, we cannot formally rule out carbon grain / PAH destruction as a source of additional carbon enhancement \citep[e.g.,][]{kastner2015,Bergin16,anderson2017,bosman2021}.  We also note that the abundance ratio is consistent with a scenario of a ``dry'' -- i.e., H$_2$O ice/gas poor -- surface  of the TW Hya disk, which also agrees with the {\em Herschel} results for the disk's cold water vapor abundance \citet{Hogerheijde11} and scattered light constraints \citep[e.g.,][]{weinberger2002,debes2013}.

\subsection{The ortho-to-para ratio of \cyc\ is consistent with a constant value of 3}
\begin{figure}
\begin{centering}
\includegraphics[width=0.45\textwidth]{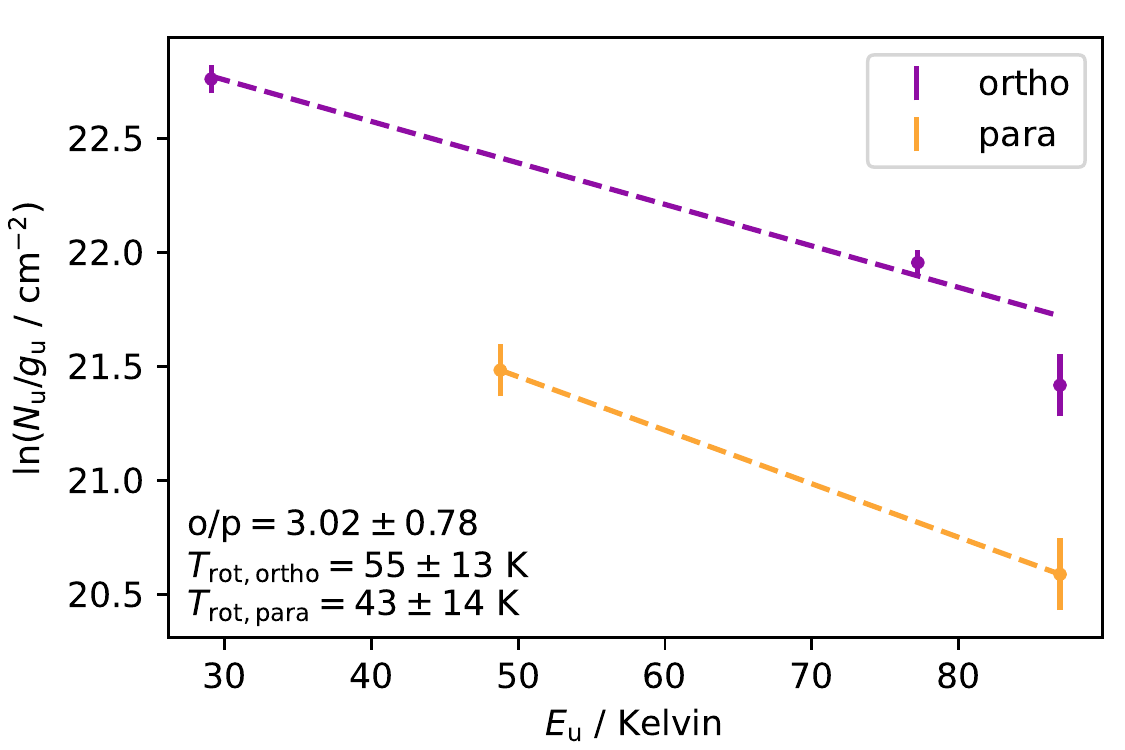}
\caption{Disk-averaged rotational diagram of the unblended lines. We find similar ortho and para rotational temperatures and a disk integrated ortho-to-para ratio consistent with 3.  \label{fig:rotdia}}
\end{centering}
\end{figure}

Via the non-LTE fitting we find that the ortho-to-para ratio across the slab can be well reproduced with a single value of 3, however we only consider values of 1 or 3 in our grid. To confirm this result, we have done an additional disk-averaged rotational diagram analysis using the unblended ortho and para transitions, which are also optically thin based on our model comparisons. Figure~\ref{fig:rotdia} shows the results of the fitting, and we find that the ortho and para lines have consistent rotational temperatures, and the derived ortho-to-para ratio from the column densities is $3.02\pm0.78$.

So where does this ortho-to-para ratio come from? Both ortho and para forms of \cyc\ mainly form via gas phase reactions \citep[e.g.,][]{Vrtilek1987,park2006}, in particular via C$_3$H$^+$ reacting with H$_2$ via radiative association.
\citet{morisawa2006} demonstrate that the ortho-to-para ratio of \cyc\ in steady state can be approximated by: 
\begin{equation}\label{eq:opr}
[o/p]_{\rm C_3H_2} = \frac{(5+3\phi)\times [o/p]_{\rm H_2} + 3\phi + 3}{(1+\phi)\times [o/p]_{\rm H_2} + \phi + 3},
\end{equation} where $\phi$ essentially reflects the ratio of the rate of formation of \cyc\ by electron dissociative recombination of C$_3$H$_3^+$ compared to reactions of C$_3$H$_3^+$  with atoms \citep[however, see also the discussion in][for additional effects]{park2006}. The parameter $\phi$ is equal to $k_{\rm DR} [n_e] / k_{\rm ion-Y} [n_{\rm Y}]$, where $k_{\rm DR}$ is the rate of dissociative recombination, $n_e$ is the number density of electrons, $k_{\rm ion-Y}$ is the reaction rate of C$_3$H$_3^+$ with species Y and $[n_{\rm Y}]$ is the number density of said species. If the ortho-to-para ratio of H$_2$ goes to 3, the dependence on $\phi$ drops out of Equation~\ref{eq:opr}. If the ortho-to-para ratio of H$_2$ is small due to a cold formation mechanism of H$_2$, then the ortho-to-para ratio of \cyc\ depends on $\phi$, going to 1 if $\phi$ is small and 3 if $\phi$ is large. Therefore, if this equation holds, obtaining the observed ortho to para ratio of 3 in \cyc\ means either H$_2$ has an ortho-to-para of 3, or $\phi$ is large.

We can estimate what $\phi$ regime \cyc\ may be in using the \citet{cleeves2018} model of a solar nebula like disk. Taking approximate values from a location of r = 50 au and $z/r$ of 0.3 au, the electron density is approximately $10^{-9}$ per H, and is largely governed by photoionization of atomic carbon. The recombination rate with electrons is approximately 10$^{-6}$ cm$^{3}$~s$^{-1}$ at 50 K \citep{Loison2017}. For the denominator, if we take reactions with C as representative, the rate is $\sim10^{-9}$ cm$^{3}$~s$^{-1}$, and the abundance is also $10^{-9}$ per H. Therefore, $\phi$ comes down to the ratio of the rate coefficients, which is $>> 1$. Therefore it is not clear whether our measured \cyc\ ortho to para ratio is shedding light on the ortho-to-para ratio of H$_2$ or if it is related to the relative rate of dissociative recombination. 

\subsection{C/O implications for forming planets}

While there are exciting prospects to use hydrocarbons to constrain C/O ratios in disks to connect with planets, we must understand what region our observations fundamentally probe. Our results find that the \cyc\ emission largely comes from the UV irradiated layer, above an $A_V$ of 1 ($z/r >0.25$). While this implies we are not directly tracing the midplane regions where planets gain most of their gas mass, it is still informative in the quest to understand what compositions planets might accrete. The bright hydrocarbon emission has been attributed to high gas-phase C/O ratios, much greater than solar \citep{Du15,kama2016,Bergin16,miotello2019}. Either the molecular layer has a carbon enhancement  (perhaps due to carbon grain destruction) or an oxygen deficit, or both. The simplest explanation for an oxygen deficit is the growth of water-ice rich grains leading to the formation of larger, midplane bound pebbles or even larger bodies as suggested by \citet{Hogerheijde11} and consistent with models of \citet{krijt2016}. A midplane ice enrichment might facilitate enhanced planet formation via an increase in solid mass, and a larger reservoir of water ice from which planets could accrete, perhaps a good scenario for forming habitable planets. 

Recently, in the HD~163296 disk, \citet{teague2019} discovered a velocity pattern in CO gas emission consistent with in-fall from the surface toward annular disk gaps. These results imply the surface gas in the disk is feeding potential planet forming regions, and that with our observations we could be tracing the same gas reservoir that might feed proto-giant planets' atmospheres \citep[see also][]{cridland2020}. Specifically in TW Hya, \citet{teague2019} found evidence of vertical motion around 90 au, near the \citet{vanBoekel16} scattered light gap. If such motions are common, this would be exciting as the layer traced by \cyc\ would be more directly related to the chemistry that sets planets' atmospheric compositions. 

Along these lines, the 20 Myr-old HR~8799 system's four outer planets have been independently chemically characterized using VLT SPHERE, and their C/O ratios radially varies \citep{lavie2017,bonnefoy2016}. The inner two planets \citep[$d$ at 24~au and $e$ at 15~au;][]{marois08,marois10} have low C/O ratios -- consistent with zero due to low C/H, while the outer two planets ($b$ at 38~au and $c$ at 68~au) have far larger C/O ratios of 0.8 -- 0.9 \citep{lavie2017}. More recently, \citet{molliere2020} reevaluated the atmospheric C/O ratios for planet $e$ and found a C/O value of 0.6, so less extreme of a jump than previously estimated. These results emphasize the need for careful treatment of non-equilibrium atmospheric effects and care in interpreting individual C/O estimates.

Interestingly, HR~8799's tentative radial increase from sub-solar/solar up to super solar values in C/O is spatially consistent with TW Hya's increase in hydrocarbon emission. In fact, the radial separation of the inner planets falls interior to the start of TW Hya's hydrocarbon ring, while the outer planets' orbits are within the radial bounds of the hydrocarbon ring. Of course there are many differences between TW Hya and the HR 8799 system, including that HR 8799's host star has a mass of 1.5 M$_\odot$ compared to TW Hya's 0.8 M$_\odot$, so such a connection is purely speculative. However, an interesting feature worth pointing out is that the younger disk IM Lup does not show the same large inner ring in hydrocarbons \citep{cleeves2018}, and instead has a more centrally peaked/flattened distribution. Rings do appear to be common among intermediate aged disks \citep{Bergin16,bergner2019}, including some of the $\gtrsim1$~Myr disks in Lupus \citep{miotello2019}. For example the 3-10 Myr \citep{bertout2007} DM Tau disk also shows a combination of an inner hydrocarbon enhancement and a secondary outer ring \citep{Bergin16}. If this qualitative observation is a truly common feature, perhaps the varied atmospheric C/O in HR 8799's planets indicate the system formed ``chemically later'' in a disk more like TW Hya than early like IM Lup. Additional observations of hydrocarbon morphologues coupled with modeling that brings together chemistry and gas-giant formation simulations are necessary to fully test this hypothesis.

\medskip

\acknowledgments 

{{\em Acknowledgements: } This paper makes use of the following ALMA data: ADS/JAO.ALMA\#2013.0.00114.S,   ADS/JAO.ALMA\#2013.0.00198.S, and \\ ADS/JAO.ALMA\#2016.0.00311.S. ALMA is a partnership of ESO (representing its member states), NSF (USA) and NINS (Japan), together with NRC (Canada), MOST and ASIAA (Taiwan), and KASI (Republic of Korea), in cooperation with the Republic of Chile. The Joint ALMA Observatory is operated by ESO, AUI/NRAO and NAOJ. The National Radio Astronomy Observatory is a facility of the National Science Foundation operated under cooperative agreement by Associated Universities, Inc. The calculations that were made for this paper were conducted on the Smithsonian Astrophysical Observatory's Hydra High Performance Cluster, for which we are grateful to have access to. L.I.C. gratefully acknowledges support from the David and Lucille Packard Foundation and the Virginia Space Grant Consortium. J.T.v.S. and M.R.H. are supported by the Dutch Astrochemistry II program of the Netherlands Organization for Scientific Research (648.000.025).  C.W.~acknowledges financial support from the University of Leeds and from the Science and Technology Facilities Council (grant numbers ST/R000549/1 and ST/T000287/1).
J.K.C. acknowledges support from the National Science Foundation Graduate Research Fellowship under Grant No. DGE 1256260 and the National Aeronautics and Space Administration FINESST grant, under Grant no. 80NSSC19K1534. V.V.G. acknowledges support from FONDECYT Iniciación 11180904. J.H. acknowledges support for this work provided by NASA through the NASA Hubble Fellowship grant \#HST-HF2-51460.001-A awarded by the Space Telescope Science Institute, which is operated by the Association of Universities for Research in Astronomy, Inc., for NASA, under contract NAS5-26555. M.K. gratefully acknowledges funding by the University of Tartu ASTRA project 2014-2020.4.01.16-0029 KOMEET, financed by the EU European Regional Development Fund.}

\bibliographystyle{aasjournal}

\appendix

Example central channels (velocity averaged for clarity) of the observed \cyc\ transitions toward TW Hya and the masks used to clean the data and generate the moment-0 maps.

\begin{figure*}[h!]
\begin{centering}
\includegraphics[width=0.87\textwidth]{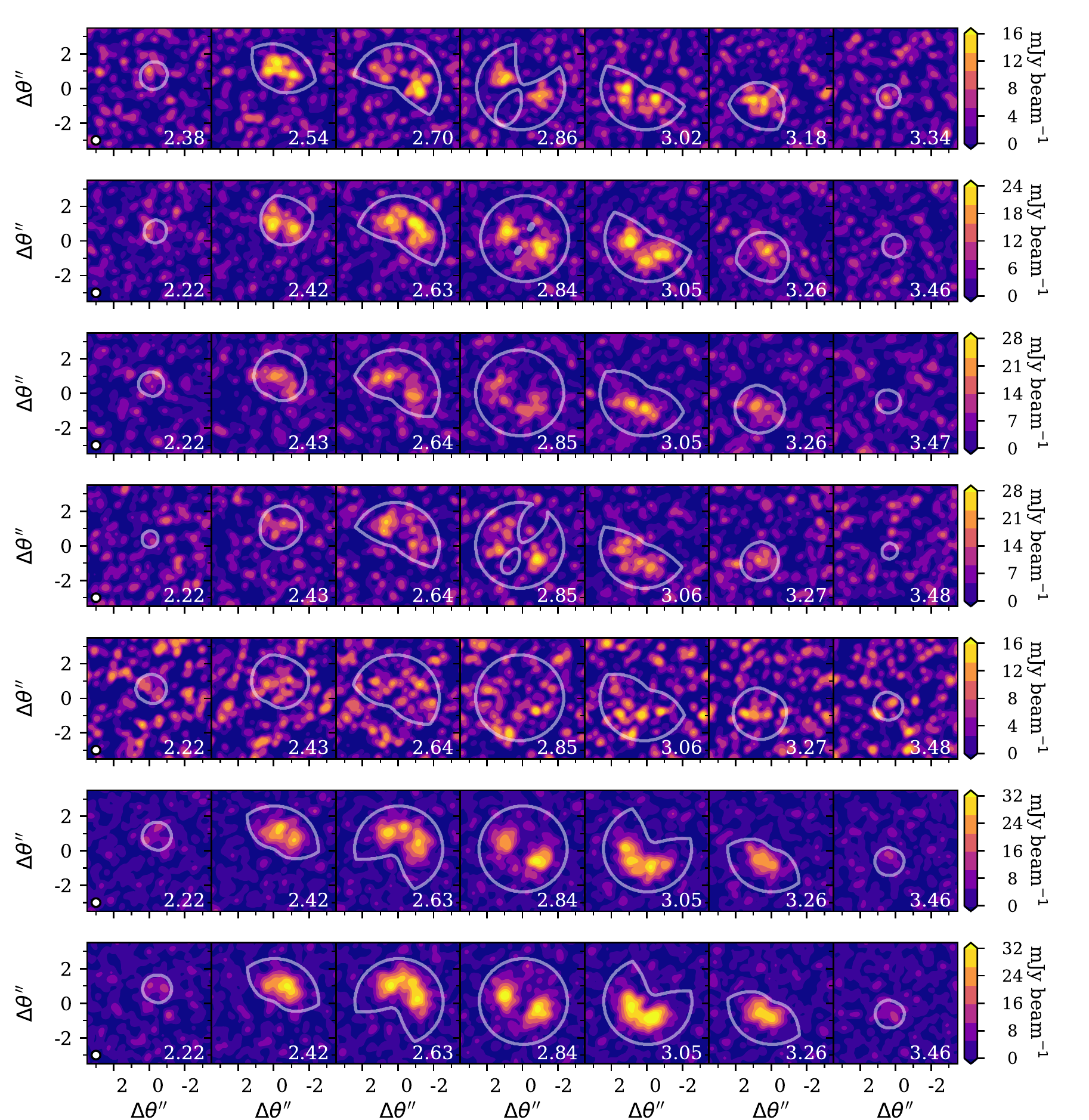}
\caption{Channel maps for the transitions of \cyc\ observed, where o = ortho, p = para, and b = blend. The contour indicates the Keplerian mask used for cleaning and for the measurements of the fluxes presented in Table~\ref{tab:linedat}. \label{fig:channels}}
\end{centering}
\end{figure*} 

\end{document}